\documentclass[twocolumn]{aastex631}

\DeclareUnicodeCharacter{02BC}{\-}
\shortauthors{Yan et al.}
\usepackage{textcomp, gensymb}
\usepackage{booktabs}
\usepackage{soul}
\usepackage{float}
\usepackage{comment}
\usepackage{array}
\usepackage{makecell}
\usepackage{amsmath}

\newcommand{\degdot}{\rlap{.}^\circ}
\newcommand{\secdot}{\rlap{.}^{\rm s}}
\newcommand{\arcsecdot}{\rlap{.}^{''}}

\begin{document}

\title{Multi-frequency Very Long Baseline Interferometry Study of Emission and Absorption in the Two-Sided Jets of NGC\,3998}

\author[0009-0003-6680-1628]{Xi Yan}
\affiliation{Xinjiang Astronomical Observatory, CAS, 150 Science 1-Street, Urumqi, Xinjiang, 830011, P.R.China}

\author[0000-0003-0721-5509]{Lang Cui}
\affiliation{Xinjiang Astronomical Observatory, CAS, 150 Science 1-Street, Urumqi, Xinjiang, 830011, P.R.China}
\affiliation{Key Laboratory of Radio Astronomy and Technology, CAS, A20 Datun Road, Chaoyang District, Beijing, 100101, P.R.China}
\affiliation{Xinjiang Key Laboratory of Radio Astrophysics, 150 Science 1-Street, Urumqi 830011, P.R.China}

\author[0000-0001-6947-5846]{Luis C. Ho}
\affiliation{Kavli Institute for Astronomy and Astrophysics, Peking University, Beijing 100871, P.R.China}
\affiliation{Department of Astronomy, School of Physics, Peking University, Beijing 100871, P.R.China}

\correspondingauthor{Lang Cui}
\email{cuilang@xao.ac.cn}

\begin{abstract}

We present the multi-frequency, multi-epoch Very Long Baseline Interferometry (VLBI) study of the two-sided jets in the low-luminosity active galactic nucleus NGC\,3998, where physical properties of the jets on parsec scales remain poorly understood. Using Very Long Baseline Array data observed at 1.4, 1.7, 2.3, and 5\,GHz, we detect symmetric twin jets aligned along the north-south direction, with a total extent of $\sim 5.3$\,pc. Notably, the position angle of the pc-scale jets differs by $26^\circ$--$30^\circ$ from that of the kpc-scale jets, suggesting the possibility of jet precession. Based on the frequency-dependent core shift and north/south jet brightness ratio, we identify the northern jet as the approaching jet and the southern jet as the counter-jet. Measurements of the radial intensity profile on both sides indicate a change in the counter-jet emission from rapid fading to a slower decline at 1.4, 1.7 and 2.3\,GHz. Spectral analysis shows that the approaching jet exhibits an optically thin spectrum, while the counter-jet is dominated by an optically thick, inverted spectrum. These findings tentatively suggest free-free absorption in NGC\,3998, which should be verified in future studies. Finally, our observations also reveal a flat-spectrum VLBI core, showing significant radio variability that is likely linked to a jet ejection event. 

\end{abstract}


\section{Introduction} \label{sec: Introduction}
The lenticular galaxy NGC\,3998 is a nearby low-luminosity active galactic nucleus \citep[LLAGN; e.g.,][]{Ho_1997ApJS..112..315H} located in the Ursa Major group. It is relatively close, at a distance of $13.7\pm1.2$\,Mpc \citep{Tonry_2001ApJ...546..681T, Cappellari_2011MNRAS.413..813C}, and hosts a supermassive black hole (SMBH) with a stellar dynamical mass of $8.1^{+2.0}_{-1.9}\times10^{8} M_{\odot}$ \citep{Walsh_2012ApJ...753...79W}. The combination of proximity and a large SMBH mass gives a scale of 1\,mas $\sim$ 0.066\,pc $\sim$ 857 Schwarzschild radii ($R_{\rm s}$), making NGC\,3998 a suitable target for studying the physical properties of the collimated relativistic jet through high-resolution very long baseline interferometry (VLBI) observations. These advantages also make NGC\,3998 one of the most promising SMBHs for future sub-millimeter VLBI imaging \citep{Ramakrishnan_2023Galax..11...15R, Johnson_2024SPIE13092E..2DJ, zhang2024accessingnewpopulationsupermassive,zineb2024advancingblackholeimaging}.

On arcminute scales, the large-scale jets in NGC\,3998 have been studied over the past few decades. Early Very Large Array (VLA) observations have revealed a compact, unresolved, bright core accompanied by faint extended emission to the south \citep{Hummel_1980A&AS...41..151H, Wrobel_1984ApJ...287...41W, Knapp_1985A&A...142....1K}. The radio core shows a flat spectrum and is variable \citep{Hummel_1984, Wrobel_1984ApJ...287...41W, Wrobel_1991AJ....101..127W,Frank_2016}. Notably, variability has also been observed in the ultraviolet \citep{Maoz_2005ApJ...625..699M, Maoz_2007MNRAS.377.1696M} and X-ray bands \citep{Ptak_2004ApJ...606..173P, Pian_2010MNRAS.401..677P, Younes_2011A&A...530A.149Y}. More recent observations have uncovered extended, S-shaped, two-sided jets, with a projected total extent of approximately 20\,kpc (or $2.6\times10^{8}\,R_{\rm s}$) along the north-south direction and an extremely low surface brightness ($S_{\rm ext}/S_{\rm tot} \approx$ 0.1--0.3) \citep{Frank_2016,Sridhar_2020A&A...634A.108S}. It is also suggested that the brightness ratio of the kpc-scale jets is relatively low, close to unity ($S_{\rm North}/S_{\rm South} \approx 0.66$--$0.76$). Furthermore, spectral analysis of the kpc-scale jets reveals a relatively homogeneous, optically thin spectral distribution on both sides ($S \propto \nu^{-0.6}$) \citep{Sridhar_2020A&A...634A.108S}. These findings suggest that relativistic Doppler effects are not strong and that the jets are nearly intrinsically symmetric on kpc scales.

However, little is known about the twin jets on arcsecond and milliarcsecond (mas) scales. From a VLBI perspective, some observations reveal a single unresolved core \citep{Baek_2019MNRAS.488.4317B}, while others suggest the presence of a jet structure extending to the north \citep{Filho_2002, Helmboldt_2007ApJ...658..203H}. Notably, no dedicated VLBI studies focusing on the NGC\,3998 jet have been published to date.

X-ray observations provide valuable insights into the nuclear physics of NGC\,3998. Different studies have indicated a nearly unabsorbed medium along the line of sight \citep[$N_{\rm H}\sim10^{20}{\rm cm^{-2}}$, e.g.,][]{Pellegrini_2000A&A...360..878P, Ptak_2004ApJ...606..173P, Pian_2010MNRAS.401..677P, Younes_2011A&A...530A.149Y, Kharb_2012AJ....143...78K, Connolly_2016MNRAS.459.3963C, Younes_2019ApJ...870...73Y, Baek_2019MNRAS.488.4317B}. Additionally, the upper limit of the narrow fluorescent Fe\,K$\alpha$ emission in NGC\,3998 has been derived to be 25\,eV \citep{Ptak_2004ApJ...606..173P}. As discussed by \citet{HO_2008ARA&A..46..475H}, the combination of low intrinsic absorption and weak Fe\,K$\alpha$ emission suggests the absence of a torus in this Type I LLAGN. On the other hand, the lack of strong reflection features, the relatively low high-energy cutoff, and multi-wavelength spectral energy distribution modeling indicate that the central accretion in NGC\,3998 is dominated by a hot accretion flow, such as advection-dominated accretion flow (ADAF) \citep[e.g.,][]{Pellegrini_2000A&A...360..878P, Ptak_2004ApJ...606..173P, Younes_2019ApJ...870...73Y}. This hot flow is truncated by a geometrically thin, optically thick accretion disk at a radius of order 100--300 $R_{\rm s}$ \citep{Ptak_2004ApJ...606..173P}.  

NGC\,3998 is also a gas-rich early-type galaxy. Optical spectral observations by {\it Hubble Space Telescope} have revealed a small, highly inclined ionized gas disk of about 100\,pc, with a position angle (PA) of $\sim84^{\circ}$ \citep[i.e., the H$\alpha$ disk;][]{Pogge_2000ApJ...532..323P,2012MNRAS.422.3208S}. Radio observations indicate the presence of a nearly polar H{\tt I} disk structure extending over 10\,kpc (PA $\sim65^{\circ}$), which is almost perpendicular to the stellar body (PA $\sim135^{\circ}$) \citep{Knapp_1985A&A...142....1K, Serra_2012MNRAS.422.1835S, Frank_2016}. Both the H$\alpha$ and H{\tt I} disks exhibit warped structures on larger scales, in a way mirroring the S-shape of the kpc-scale jets \citep{Frank_2016}. These warped gas disks suggest that the accreting material in NGC\,3998 may originate from a minor merger with a gas-rich small galaxy.

In this paper, we present the first multi-frequency, multi-epoch VLBI study of the two-sided jets in NGC\,3998. Section~\ref{sec: Observations and Data Reduction} details the observations and data reduction. Section~\ref{sec: Results and Discussion} shows the results and their corresponding discussion. 
Our main findings are summarized in Section~\ref{sec: summary}. Throughout this paper, the spectral index, $\alpha$, is defined as $S\propto\nu^{+\alpha}$, where $S$ and $\nu$ are flux density and frequency, respectively.

\begin{deluxetable*}{ccccccccccc}[hbtp!]
\tabletypesize{\footnotesize}
\tablecaption{Summary of NGC\,3998 Observations \label{tab: NGC3998_summary_of_observation}}
\tablehead{ \colhead{P.C.} & \colhead{$\nu$} & \colhead{Date} & \colhead{Array} & \colhead{Pol.} & \colhead{B.W.} & \colhead{$T_{\rm int}$} & \colhead{$\theta_{\rm maj} \times \theta_{\rm min}$, PA} & \colhead{ $I_{\rm peak}$} & \colhead{$I_{\rm rms}$} & \colhead{$S_{\rm tot}$}\\
& (GHz) & (yyyy/mm/dd) & & & (MHz) & (min) & (mas $\times$ mas, $\circ$) & (mJy/beam) & (mJy/beam) & (mJy)\\
(1) & (2) & (3) & (4) & (5) & (6) & (7) & (8) & (9) & (10)& (11)
}
\startdata
BJ037   & 1.4  & 2001/02/07  & VLBA & LCP & 64 & $\sim130$ & $10.73\times6.58, -1.1 $ & 87.8 & 0.11 & $111.8\pm11.2$\\
        & 1.7  & 2001/02/07  & VLBA & LCP & 64 & $\sim130$ & $8.72\times5.34, -3.2$  & 77.6 & 0.11 & $101.3\pm10.1$\\
        & 2.3  & 2001/02/11  & VLBA & RCP & 64 & $\sim130$ & $6.19\times3.70, -4.5$  & 82.4 & 0.12 & $112.5\pm11.3$\\
        & 5.0  & 2001/02/11  & VLBA & LCP & 64 & $\sim130$ & $2.90\times1.72, -3.3$  & 80.7 & 0.10 & $100.7\pm10.1$\\
BN021D  & 5.0  & 2003/01/23  & VLBA, -BR & LCP & 32 & $\sim 50$ & $5.66\times1.41, -2.4$  & 99.0 & 0.18 & $116.4\pm11.6$\\
        & 5.0  & 2004/01/19  & VLBA & LCP & 32  & $\sim50$ &  $4.20\times1.45, -4.3$  & 102.7 & 0.16 & $122.4\pm12.2$\\
        & 5.0  & 2004/03/19  & VLBA & LCP & 32  & $\sim50$ &  $4.16\times1.44, -4.3$ & 101.6 & 0.15 & $119.8\pm12.0$\\
        & 5.0  & 2004/05/22  & VLBA & LCP & 32  & $\sim50$ &  $4.21\times1.44, -5.0$ & 105.0 & 0.14 & $123.4\pm12.3$\\
        & 5.0  & 2004/07/19  & VLBA & LCP & 32  & $\sim50$ &  $4.23\times1.42, -4.0$ & 114.3 & 0.18 & $132.0\pm13.2$\\
        & 5.0  & 2004/09/16  & VLBA, -KP & LCP & 32 & $\sim50$ & $4.11\times1.38, -4.6$ & 114.9 & 0.17 & $133.6\pm13.4$\\
        & 5.0  & 2004/11/20  & VLBA & LCP & 32  & $\sim50$ &  $4.38\times1.44, -6.9$ & 126.9 & 0.16 & $142.4\pm14.2$\\
\enddata
\tablecomments{
Column\,(1): project code.
Column\,(2): frequency. 
Column\,(3): epoch. 
Column\,(4): participating stations. The minus denotes that the station did not participate due to some issues.
Column\,(5): polarization. 
Column\,(6): bandwidth.
Column\,(7): on-source integration time.
Column\,(8): Full Width at Half Maximum (FWHM) of the synthesized beam's major and minor axes, along with the position angle (PA) of the major axis.
Columns\,(9)-(11): peak intensity, rms noise level ($\sigma$), and total flux density of the CLEAN image.
}
\end{deluxetable*}

\section{Observations and Data Reduction} \label{sec: Observations and Data Reduction}
This study is based on archival Very Long Baseline Array (VLBA) data observed at 1.4, 1.7, 2.3, and 5\,GHz, as summarized in Table~\ref{tab: NGC3998_summary_of_observation}.

\subsection{BJ037} \label{sec: BJ037}
The BJ037 experiment was conducted at 1.4/1.7\,GHz on 2001 February 07 and at 2.3/5\,GHz on 2001 February 11. The data were recorded in either left circular polarization (LCP) or right circular polarization (RCP), with a total bandwidth of 64\,MHz. We note that these observations involved fast switching between NGC\,3998 and the calibrator J1148+5254 (separated by 2$\degdot$9 on the sky), with each frequency alternating in turn every $\sim$ 5 minutes. This resulted in an on-source integration time of about 130 minutes for each frequency (see Table~\ref{tab: NGC3998_summary_of_observation}).

The data were calibrated following standard procedures using the Astronomical Image Processing System \citep[AIPS;][]{Greisen2003}, developed by the National Radio Astronomy Observatory (NRAO). After removing the digital sampler biases by examining the auto-correlation spectra, we performed phase calibration as follows. We first corrected for the parallactic angle, ionospheric fluctuations, earth orientation parameters, and instrumental delays. Then, the global fringe fitting was run on the target and calibrators to remove the residual phases, delays, and rates assuming a point source model. Using the antenna gain curves and system temperatures, we performed priori amplitude calibration. Finally, bandpass solutions were derived from scans of bright fringe-finding sources. We averaged the calibrated data over frequency and load them into the Caltech DIFMAP package \citep{Shepherd_Difmap_1997ASPC..125...77S} for self-calibration and imaging.

\begin{figure*}[htbp!]
\begin{center}
    \includegraphics[width=0.41\linewidth]{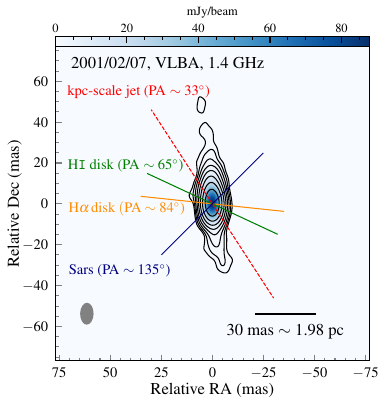}
    \includegraphics[width=0.40\linewidth]{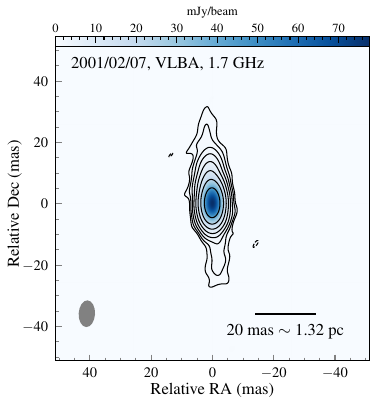}
    \includegraphics[width=0.41\linewidth]{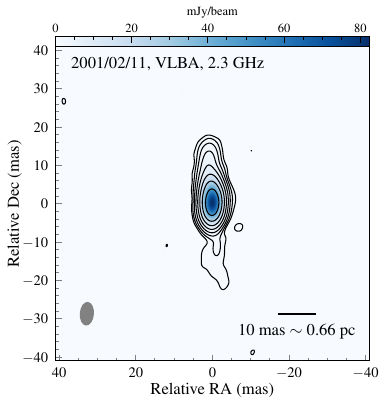}
    \includegraphics[width=0.41\linewidth]{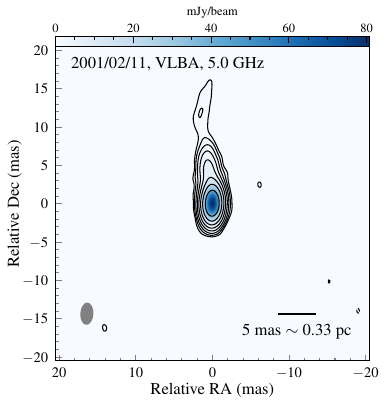}
\caption{Naturally weighted CLEAN images of NGC\,3998 at 1.4, 1.7, 2.3, and 5.0\,GHz. In the 1.4\,GHz image (top-left panel), the position angles of the nearly edge-on gas disks (i.e., the H$\alpha$ and H{\tt I} disks), the stellar body, and the kpc-scale jet are indicated. The synthesized beam is shown in the bottom-left corner of each image. Contours are plotted at $-1$, $1$, $2$, $4$, ..., times the $3\sigma$ noise level of each image (see Table~\ref{tab: NGC3998_summary_of_observation}), increasing by a factor of 2.
\label{fig: NGC3998_CLEAN_images}}
\end{center}
\end{figure*}

We also conducted another round of data calibration for the astrometric purpose. 
The calibration steps were similar to those described above, except that global fringe fitting was run only on the target NGC\,3998, using its source model created by previous imaging process at each frequency (see Figure~\ref{fig: NGC3998_CLEAN_images}). The residual phase, delay, and rate solutions were then interpolated to the consecutive scans of the calibrator J1148+5254. We did not use J1148+5254 as the phase-referencing source due to the following reasons: 1) its total flux density at 1.4 and 1.7\,GHz is relatively low ($\sim80$\,mJy; see Table~\ref{tab: Flux density of J1148+5254} and Figure~\ref{fig: J1148+5254_spectra} in Appendix~\ref{Appendix: Calibrator J1148+5254}); 2) it exhibits a noticeable ``core + knot" structure at 1.4 and 1.7\,GHz (see Figure~\ref{fig: J1148+5254_PRed_CLEAN}), resulting in significant phase variations as a function of $uv$ distance (between $-100^{\circ}$ and $100^{\circ}$); and 3) the unique structure of J1148+5254 makes it more suitable for identifying the precise astrometric positions of the core and knot from its phase-referenced images. In contrast, NGC\,3998 shows a higher flux density at 1.4 and 1.7\,GHz (100--110\,mJy, see Table~\ref{tab: NGC3998_summary_of_observation}) compared to J1148+5254  and has much more stable phase in the visibility domain due to its core-dominated nature. After the calibration was done, we read the data into DIFMAP and successfully obtained phase-referenced images of J1148+5254, from which we determined the astrometric position of J1148+5254 relative to NGC\,3998.

\subsection{BN021D}
In the BN021D project, NGC\,3998 was observed at 5\,GHz over seven epochs between 2003 and 2004. However, the on-source integration time for each observation was relatively short, approximately 50 minutes (see Table~\ref{tab: NGC3998_summary_of_observation}). The data reduction process followed similar procedures to those applied to the BJ037 dataset, including amplitude and bandpass calibration, as well as global fringe fitting for each source (without phase referencing). In this study, the multi-epoch 5\,GHz data were utilized to analyze the variability of NGC\,3998.

\section{Results and Discussion} \label{sec: Results and Discussion}
\subsection{Jet morphology}
The multi-frequency VLBI images of NGC\,3998 obtained from project BJ037 are presented in Figure~\ref{fig: NGC3998_CLEAN_images}, while the multi-epoch 5\,GHz images from project BN021D are shown in Appendix~\ref{Appendix: The multi-epoch 5 GHz CLEAN images}. The image parameters are summarized in Table~\ref{tab: NGC3998_summary_of_observation}. At 1.4 and 1.7\,GHz, the two-sided jets are clearly detected along the north-south direction, with a total extent of $\sim 80$\,mas, corresponding to $\sim 5.3$\,pc. The position angles of the northern and southern jets are $3^\circ$--$7^\circ$ and $-174^\circ$ -- $-177^\circ$, respectively. 

In the 1.4\,GHz image of Figure~\ref{fig: NGC3998_CLEAN_images}, we show the position angles of various components of this galaxy, including the H$\alpha$ disk, the H{\tt I} disk, and the kpc-scale jet. Notably, the pc-scale jet (PA $\approx 3^{\circ}$–$7^{\circ}$, observed in February 2001) is misaligned with the kpc-scale jets (PA $\sim 33^{\circ}$, observed in March 2015; \citealt{Sridhar_2020A&A...634A.108S}) by about $26^\circ$–$30^\circ$. One possible explanation for this difference could be jet precession, as similar to M\,87, where a precessing jet has been suggested based on variations in the jet position angle \citep{Cui_2023Natur.621..711C}. The possibility of jet precession in NGC\,3998 has been explored by \citet{Frank_2016}, who based their analysis on observed warped gas disks and their connection to the kpc-scale S-shaped radio lobes. According to these authors, the S-shaped radio morphology could result from the jet axis adapting to the changing angular momentum of the accreting gas. Apparently, long-term monitoring of the NGC\,3998 jets at both mas and arcsecond scales will be crucial to confirm the precession scenario.

The pc-scale jet appears to be more aligned with the apparent rotation axis of the H$\alpha$/H{\tt I} disks compared to the kpc-scale jet (see Figure~\ref{fig: NGC3998_CLEAN_images}). However, as suggested by, for example, \citet{Wu_2022ApJ...941...95W}, double-lobed radio jets in giant elliptical galaxies generally do not show a preferential alignment with the minor axes of the gas disks \citep[see also the review by][]{Kormendy_2013ARA&A..51..511K}. Thus, the observed alignment in NGC\,3998 is likely fortuitous, possibly resulting from the influence of stellar torques that warp the inner gas disks \citep{Frank_2016}.

\begin{figure*}[htbp!]
\begin{center}
    \includegraphics[width=1.0\linewidth]{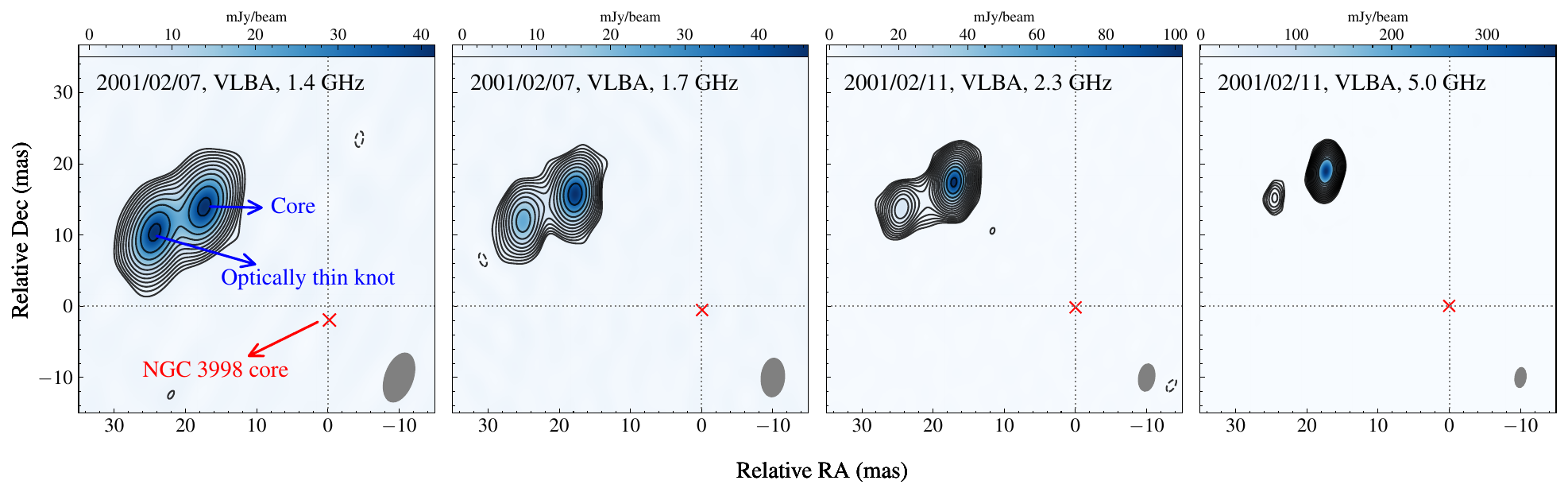}
\caption{Phase-referenced images of J1148+5254 observed at 1.4, 1.7, 2.3, and 5.0\,GHz. A uniform weighting scheme was used for the 1.4, 1.7, and 2.3\,GHz images, while a natural weighting scheme was applied to the 5.0\,GHz image to enhance signal-to-noise ratio (S/N) of the weak knot. Contours start at the 3$\sigma$ noise level of each image (i.e., 0.4, 0.4, 0.3, 0.4\,mJy/beam, respectively) and increase by a factor of $\sqrt{2}$. The synthesized beam is shown in the bottom-right corner of each image, and the red cross marks the derived position of NGC\,3998 core. The spectral properties of J1148+5254 are provided in Appendix~\ref{Appendix: Calibrator J1148+5254}.
\label{fig: J1148+5254_PRed_CLEAN}}
\end{center}
\end{figure*}

\begin{deluxetable*}{ccccccccc}[htbp!]
\tabletypesize{\scriptsize}
\tablecaption{Model-fitted Parameters Used to Derive the Core Shifts of NGC\,3998 
\label{tab: NGC3998_core_shift_parameters}}
\tablehead{
\colhead{$\nu$} & \multicolumn{2}{c}{J1148+5254} & \multicolumn{2}{c}{NGC\,3998} & \colhead{$\Delta {\rm Dec}$} & \colhead{$\Delta {\rm RA}$} & \colhead{$\Delta r_{\rm core, Dec}$} & \colhead{$\Delta r_{\rm core, RA}$ }\\
\cmidrule(r){2-3}  \cmidrule(r){4-5} 
 & $r$ & PA  & $r$ & PA \\
 (GHz) & (mas) & ($\circ$) & (mas) & ($\circ$) & (mas) & (mas) & (mas) & (mas)  \\
(1) & (2)& (3)& (4)& (5)& (6)& (7) & (8)& (9)}
\startdata
1.4 & $26.42\pm0.10$ & $66.9\pm0.2$ & $1.94\pm0.83$ & $-173.2\pm5.7$ & $12.29\pm0.94$ & $24.53\pm0.94$ & $2.88\pm0.94$ & $-0.07\pm0.94$  \\
1.7 & $27.67\pm0.12$ & $63.9\pm0.2$ & $0.52\pm0.68$ & $-173.3\pm5.7$ & $12.69\pm0.64$ & $24.91\pm0.65$& $2.48\pm0.64$ & $-0.45\pm0.65$    \\
2.3 & $27.87\pm0.20$ & $60.6\pm0.4$ & $0.15\pm0.48$ & $-172.5\pm5.7$ & $13.83\pm0.37$ & $24.30\pm0.39$& $1.34\pm0.37$ & $0.16\pm0.39$     \\
5.0 & $28.82\pm0.18$ & $58.1\pm0.4$ & $0.06\pm0.22$ & $7.32\pm5.7$ & $15.17\pm0.14$ & $24.46\pm0.18$ & $0\pm0.14$ & $0\pm0.18$   \\
\enddata
\tablecomments{
Column\,(1): frequency.
Columns\,(2)-(3): distance and PA of the optically thin knot of J1148+5254 relative to the image center.
Columns\,(4)-(5): distance and PA of NGC\,3998 core relative to the image center.
Columns\,(6)-(7): relative distance between the NGC\,3998 core and the optically thin knot of J1148+5254, along Dec and RA directions, respectively.
Columns\,(8)-(9): measured core positions relative to the 5\,GHz core along Dec and RA directions for NGC\,3998, calculated from Columns\,(6) and (7), respectively. The positional uncertainties are adopted from Table~\ref{tab: NGC3998_core_shift_error_budget}.
}
\end{deluxetable*}

\begin{deluxetable*}{ccccccc}
\tablecaption{Error Budgets for the Astrometric Observations ($\mu$as) \label{tab: NGC3998_core_shift_error_budget}
}
\tablehead{ \colhead{$\nu$ (GHz)} & \colhead{1.4} & \colhead{1.7} & \colhead{2.3} & \colhead{5.0} 
}
\startdata
Thermal noise error & 68 & 48 & 16 & 3 \\
Ionosphere & 930 & 630 & 344 & 73 \\
Troposphere & 68 & 68 & 68 & 68 \\
Earth orientation & 5 & 5 & 5 & 5 \\
Antenna position & 3 & 3 & 3 & 3\\
Priori source coordinates & 5 & 5 & 5 & 5\\
Position error of J1148+5254's knot (Dec/RA) & 37/88 & 52/108 & 97/173 & 96/154 \\
Position error of NGC\,3998 core (Dec/RA) & 28/3 & 29/3 & 20/3 & 9/1 \\
\hline
Root-sum-squared (Dec/RA)  & 936/939 & 638/645 & 365/391 & 139/184  \\
\enddata
\tablecomments{
The errors are estimated for both Dec and RA directions. Specifically, the errors include contributions from thermal noise in each phase-referenced image, residual ionospheric and tropospheric effects, Earth orientation parameter inaccuracies, antenna and source position errors, position uncertainties of the J1148+5254 knot and NGC\,3998 core (see text for details). Note that the positional uncertainty of NGC\,3998 core is likely underestimated (see the text). Furthermore, }the ionospheric error may also be underestimated due to the uncertainties arising from the non-simultaneous observations at 1.4/1.7\,GHz and 2.3/5.0\,GHz.
\end{deluxetable*}

\vspace{-1.8cm}
\subsection{Astrometry of the core} \label{sec: Astrometry of the core}
As described in Section~\ref{sec: BJ037}, we determined the astrometric position of J1148+5254 relative to NGC\,3998. This provides a basis for measuring the frequency-dependent position of the core in NGC\,3998 (i.e., core shift), which is attributed to the synchrotron self-absorption (SSA) effect \citep[e.g.,][]{Konigl_1981ApJ...243..700K,Lobanov_1998A&A...330...79L}. 

Figure~\ref{fig: J1148+5254_PRed_CLEAN} presents phase-referenced images of J1148+5254 observed at 1.4, 1.7, 2.3, and 5\,GHz\footnote{It should be noted that the phase-referenced position of J1148+5254 is offset by as much as 20\,mas from the phase-tracking center (see Table~\ref{tab: NGC3998_core_shift_parameters}). However, given that the VLBI-scale source coordinates for both NGC\,3998 and J1148+5254 have been relatively well-constrained by previous geodetic and calibrator surveys --- typically with accuracies of a few mas or better \citep[e.g.,][]{Petrov_2011AJ....142...89P} --- this discrepancy is unexpected. This is possibly because the nominal source coordinates for NGC\,3998, used in the BJ037 observations, are not sufficiently accurate. We note that the J2000 position for NGC\,3998 used in the observations ($11^{\rm h}57^{\rm m}56\secdot1350$, $55^{\circ}27^{'}12\arcsecdot950$) differ by 25.5–28.0\,mas from those reported in the \href{https://astrogeo.org/vlbi_images/}{Astrogeo dataset} ($11^{\rm h}57^{\rm m}56\secdot1333$, $55^{\circ}27^{'}12\arcsecdot922$).}. Notably, this source exhibits a relatively simple structure: a compact, optically thick core ($\alpha = 1.8 \pm 0.1$), accompanied by a well-isolated, optically thin knot ($\alpha = -1.6 \pm 0.1$) located to the southeast\footnote{Appendix~\ref{Appendix: Calibrator J1148+5254} includes the spectral data of J1148+5254.}.
The position of the optically thin component is usually assumed to be frequency-independent, making it a reliable reference position \citep[e.g.,][]{Kovalev_2008A&A...483..759K,hada_evidence_2013,Fromm_2013A&A...557A.105F}. Using the {\tt MODELFIT} subroutine in DIFMAP, we determined the positions of the optically thin knot in J1148+5254 relative to the image center at each frequency. 
Subsequently, we identified the core position in each self-calibrated image of NGC\,3998 (Figure~\ref{fig: NGC3998_CLEAN_images}) by fitting multiple circular Gaussian functions to the visibility data. The relative distance between the knot in J1148+5254 and the NGC\,3998 core was then calculated (see Figure~\ref{fig: J1148+5254_PRed_CLEAN} and Table~\ref{tab: NGC3998_core_shift_parameters}). Finally, we determined the relative positions of the core observed at different frequencies for NGC\,3998, i.e., the core shifts (Table~\ref{tab: NGC3998_core_shift_parameters}).

The positional uncertainties in our astrometric observations were estimated following the methods outlined in \citet{Hada_2011Natur.477..185H} and \citet{haga2015ApJ80715H}. The thermal noise error was estimated as the ratio of major-axis size of the synthesized beam to the S/N of each phase-referenced image (Figure~\ref{fig: J1148+5254_PRed_CLEAN}). The ionospheric residuals were estimated on basis of a total electron content of $5 \times 10^{17}\,{\rm m^{-2}}$ (determined from Global Positioning System measurements), adopting an uncertainty of 25\% and a low antenna elevation angle of $4^{\circ}$. The tropospheric residuals were estimated assuming a 3\,cm zenith excess path error for a zenith angle of $50^{\circ}$ \citep{Beasley_1995ASPC...82..327B}. Geometric errors, including Earth orientation parameter errors, antenna position errors, and source position errors, were estimated based on the source declination \citep{Pradel_2006A&A...452.1099P}. 
Positional uncertainty of the knot in J1148+5254 --- primarily influenced by the quality of the phase-referenced image \citep[see relevant discussion in][]{Hada_2011Natur.477..185H} --- was estimated based on its radial distance and S/N \citep{Fomalont_1999ASPC..180..301F}. Positional uncertainty of the NGC\,3998 core was also estimated based on the S/N. However, deriving the positional error purely based on the S/N does not account for systematic errors associated with calibration and image processes. Therefore, we emphasize that the positional uncertainty of NGC\,3998 core could be underestimated. A summary of the estimated error budgets is provided in Table~\ref{tab: NGC3998_core_shift_error_budget}.

\begin{figure}[t!]
\begin{center}
        \centering
        \includegraphics[width=1\linewidth]{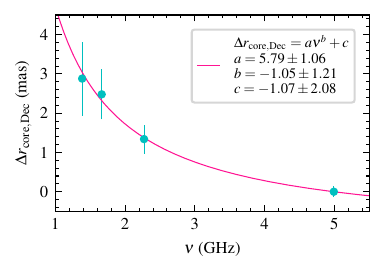}
    \caption{Core position as a function of frequency. The pink line represents the fitted power-law function to these measurements, with the function and the derived parameters noted in the upper-right corner.
    \label{fig: NGC3998_core_shift}}
\end{center}
\end{figure}

\begin{figure*}[htbp!]
\begin{center}
    \includegraphics[width=1\linewidth]{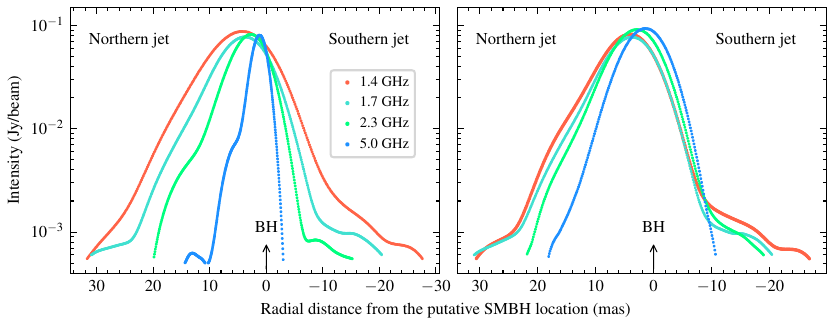}
\caption{Jet intensity profile as a function of radial distance from the putative location of the central SMBH, measured at 1.4, 1.7, 2.3, and 5.0\,GHz. All measurements are taken with S/Ns $\geq$ 5. Left: Profiles derived from the self-calibrated images shown in Figure~\ref{fig: NGC3998_CLEAN_images}. Right: Profiles derived from images restored with uniform map and beam sizes (see details in the text).
\label{fig: NGC3998_intensity_profile}}
\end{center}
\end{figure*}

As shown in Columns (8) and (9) of Table~\ref{tab: NGC3998_core_shift_parameters}, the core position shifts predominantly northward with decreasing frequency, suggesting that the southern jet is the counter-jet. In contrast, the shifts along the RA direction are negligible. Figure~\ref{fig: NGC3998_core_shift} presents the frequency-dependent core shifts in Dec direction. A power-law fit to these measurements yields $\Delta r_{\rm core, Dec} \propto \nu^{-1.05\pm1.21}$. We emphasize that the fitted parameters have substantial uncertainties, primarily due to the limited number of measurements. With this limitation in mind, we imply that this result appears to broadly align with the measured core shifts in other LLAGN \citep[e.g.,][]{Hada_2011Natur.477..185H,haga2015ApJ80715H,park_2021ApJ...909...76P,boccardi_2021A&A...647A..67B} and be consistent with theoretical predictions \citep[e.g.,][]{Konigl_1981ApJ...243..700K}. The SMBH may be located approximately 1.07\,mas upstream of the 5\,GHz core, corresponding to $\sim$900 $R_{\rm s}$ in projection. In future work, we plan to further investigate the multi-frequency properties of the VLBI core using new observations at higher frequencies. In the following sections, we utilize the measured core shifts to analyze the radial intensity profiles of the twin jets and examine their spectral properties across different frequencies.

The measured core shift for NGC\,3998 appears to be relatively large, i.e., $r_{\rm 1.4/5\,GHz} = 2.88\pm0.94$\,mas. Similar large core shifts have been reported in other LLAGN. For instance, \citet{haga2015ApJ80715H} measured a core shift of $r_{\rm 1.4/5\,GHz} = 3.74\pm2.06$\,mas for the approaching jet and $r_{\rm 1.4/5\,GHz} = 2.77\pm2.05$\,mas for the receding jet in NGC\,4261. In NGC\,315, \citet{park_2021ApJ...909...76P} derived $r_{\rm 1.5/5\,GHz} = 3.61\pm0.51$\,mas, and \citet{boccardi_2021A&A...647A..67B} independently determined $r_{\rm 1.4/5\,GHz} = 4.0\pm0.60$\,mas \citep[see also][]{Ricci_2022cosp...44.2059R,Ricci_2025A&A...693A.172R}.


\subsection{Radial intensity profile of the two-sided jets} \label{sec: Radial intensity profile}
To analyze the longitudinal intensity distribution of jet as a function of distance, we determined the jet ridge line by identifying the brightest pixel in each transversely sliced intensity profile. Measurements with S/Ns $<5$ were excluded. The results are shown in the left panel of Figure~\ref{fig: NGC3998_intensity_profile}, with the SMBH shifted to the coordinate origin based on the measured core shifts described in Section~\ref{sec: Astrometry of the core}. An intriguing finding is the evolution of the intensity profile for the southern jet (i.e., the counter-jet). Within the inner $\sim$10\,mas region from the SMBH, the jet intensity fades rapidly with increasing distance, while beyond $\sim$10\,mas, the intensity decreases more slowly. This phenomenon is more pronounced at 1.7 and 2.3\,GHz, where a small dip appears in each intensity profile. In contrast, the northern jet fades more gradually and does not exhibit the significant variation trend observed in the southern jet.

To compare the radial intensity profiles across different frequencies, we restored all images using a circular beam with a size of $\theta_{\rm eq, 1.4\,GHz} \approx 8.4$\,mas ($\theta_{\rm eq} = \sqrt{\theta_{\rm maj} \times \theta_{\rm min}}$ is the average equivalent beam size, where $\theta_{\rm maj}$ and $\theta_{\rm min}$ are provided in Table~\ref{tab: NGC3998_summary_of_observation}). We used the same map size for all images and then extracted the intensity distribution of the jet in each image. As shown in the right panel of Figure~\ref{fig: NGC3998_intensity_profile}, the intensity variation trend in the southern jet is more evident at 1.4, 1.7, and 2.3\,GHz. Another notable finding is the spectral behavior of the jets: the northern jet exhibits an optically thin spectrum, while the southern jet is dominated by an optically thick, inverted spectrum. Interestingly, beyond a distance of $\sim$16\,mas, the spectrum of the southern jet transitions to being optically thin. Absorption of the southern jet emission in the inner may explain the evolution in its radial intensity profile and its inverted spectra.


\subsection{Spectral index maps} \label{sec: Spectral index maps}
Using the two-epoch data from project BJ037 (separated by four days, see Table~\ref{tab: NGC3998_summary_of_observation}), we generated spectral index maps (SIMs) for frequency pairs 1.4/2.3\,GHz, 1.7/2.3\,GHz, and 2.3/5.0\,GHz through the following steps:

\begin{itemize}
    \item[(1)] \textit{Matching $uv$-coverage:} to ensure comparable $uv$-coverage between adjacent frequencies, we excluded long-baseline data at higher frequency and short-baseline data at lower frequency \citep[e.g.,][]{Hovatta_2014AJ....147..143H,Ro_2023A&A...673A.159R}. The resulting $uv$-ranges are as follows:
   \begin{align*}
   \text{1.4/2.3\,GHz}: &\ 1.4 \text{--} 40\,{\rm M}\lambda \\
   \text{1.7/2.3\,GHz}: &\ 1.6 \text{--} 48\,{\rm M}\lambda \\
   \text{2.3/5.0\,GHz}: &\ 3.0 \text{--} 65\,{\rm M}\lambda \\
   \end{align*}

    \vspace{-0.8cm}
    \item[(2)] \textit{Image preparation:} using the clipped data, we obtained new self-calibrated images. To ensure consistency, the same map size and circular restoring beam were used for each frequency pair. The beam size was set as the average equivalent beam size ($\theta_{\rm eq}$) of the lower-frequency image.

    \item[(3)] \textit{Spectral index calculation:} using the measured core shifts (Section~\ref{sec: Astrometry of the core}), the images were properly aligned. Then, the SIMs were generated by calculating the spectral index at each pixel using the formula:
    \begin{equation} \label{eq: alpha_ij}
          \alpha_{ij} = \frac{\text{log}(I_{\nu_2,ij}/I_{\nu_1,ij})}{\text{log}(\nu_2/\nu_1)}
   \end{equation}
    where $I_{\nu,ij}$ represents the intensity at pixel $(i, j)$ for the observing frequency $\nu$. Pixels with $I_{\nu,ij} \leq 3\sqrt{\sigma_{\rm rms,\nu}^{2} + (1.5\sigma_{\rm rms,\nu})^{2}} \approx 5.4\sigma_{\rm rms,\nu}$ were blanked, where $\sigma_{\rm rms,\nu}$ is the rms noise level of the image \citep{Hovatta_2014AJ....147..143H}.
    
    \item[(4)] \textit{Error estimation:} the uncertainty in $\alpha_{ij}$ was estimated using standard error propagation \citep[e.g.,][]{Kim_2014JKAS...47..195K}:
    \begin{equation} \label{eq: sigma_alpha_ij}
        \sigma_{\alpha_{ij}} = \frac{1}{\text{log}(\nu_2/\nu_1)} \times \sqrt{\left(\frac{\sigma_{I_{\nu_1,ij}}}{I_{\nu_1,ij}}\right)^{2} + \left(\frac{\sigma_{I_{\nu_2,ij}}}{I_{\nu_2,ij}}\right)^{2}}
    \end{equation}
    where $\sigma_{I_{\nu,ij}} = \delta_{\nu}\,I_{\nu,ij} + \sigma_{\rm rms,\nu}$ represents the uncertainty in intensity $I_{\nu,ij}$, with $\delta_{\nu} \approx 0.1$ denoting the systematic amplitude calibration error.

    \item[(5)] \textit{Spectral distribution:} the spectral distribution along the jet was obtained by calculating the weighted average spectral indices perpendicular to the jet axis as follows:  
    \begin{equation}
        \bar{\alpha}_j = \frac{\sum_{i=1}^{n} (\alpha_{ij} \sigma_{\alpha_{ij}}^{-2})}{\sum_{i=1}^{n} \sigma_{\alpha_{ij}}^{-2}}, \quad j = 1, 2, 3, \ldots
    \end{equation}
    where $\bar{\alpha}_j$ denotes the weighted average spectral index for $n$ pixels along the $j$-th segment of the transverse jet, with $\alpha_{ij}$ and $\sigma_{\alpha_{ij}}$ given by Eqs.\,\ref{eq: alpha_ij} and \ref{eq: sigma_alpha_ij}, respectively. The uncertainty of $\bar{\alpha}_j$ is estimated by $(n/{\sum_{i=1}^{n} \sigma_{\alpha_{ij}}^{-2}})^{1/2}$ \citep[e.g.,][]{Park_2019ApJ...871..257P,Ro_2023A&A...673A.159R}.
    
\end{itemize}

\begin{figure*}[htbp!]
\begin{center}
    \includegraphics[width=0.34\linewidth]{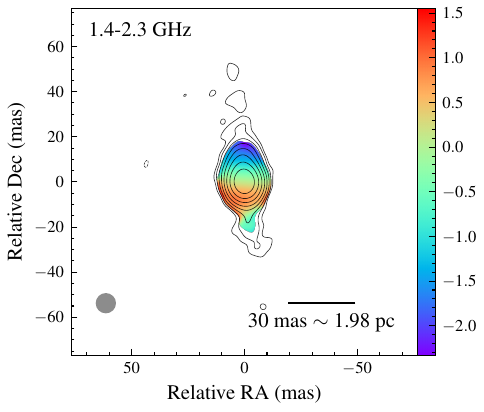}
    \includegraphics[width=0.31\linewidth]{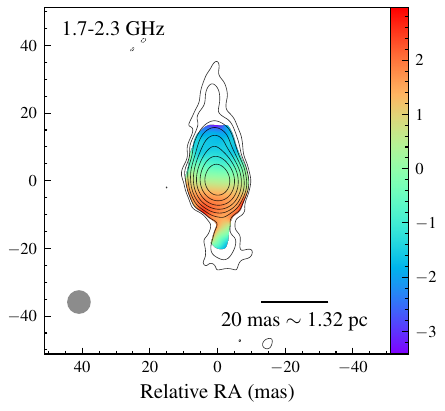}
    \includegraphics[width=0.32\linewidth]{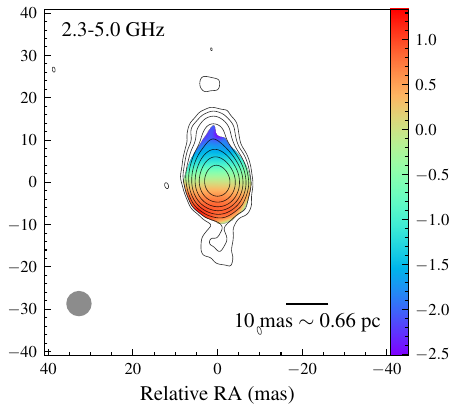}
    \includegraphics[width=0.34\linewidth]{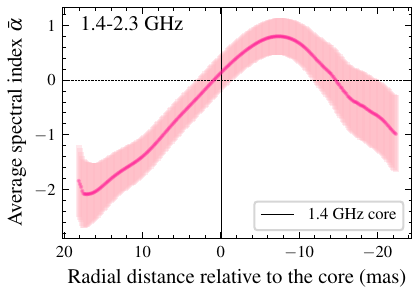}
    \includegraphics[width=0.32\linewidth]{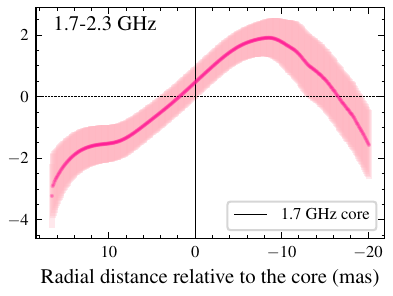}   
    \includegraphics[width=0.32\linewidth]{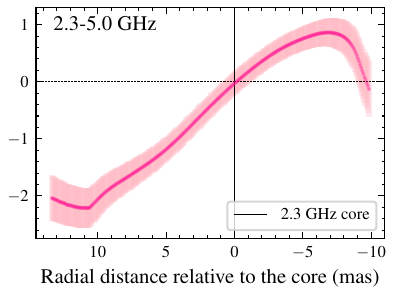}
\caption{Top: SIMs for different frequency pairs, overlaid on the total intensity contours of the lower-frequency image. The restoring beam is plotted in the bottom-left corner of each image. Bottom: the weighted average spectral index as a function of radial distance from the radio core for different frequency pairs. The peak values of the inverted spectra in the southern jet are $\bar{\alpha}_{\rm 1.4/2.3\,GHz} = 0.80\pm0.33$, $\bar{\alpha}_{\rm 1.7/2.3\,GHz} = 1.92\pm0.62$, and $\bar{\alpha}_{\rm 2.3/5.0\,GHz} = 0.86\pm0.25$. 
We note that the measured spectral indices are sparse at the two edges of the jet, likely due to limited sampling on short baselines after matching the $uv$-coverage. As a result, the significantly steep spectra observed at the edges of jet may be spurious.}
\label{fig: NGC3998_SIMs}
\end{center}
\end{figure*}

\begin{figure*}[htbp!]
\begin{center}
    \includegraphics[width=0.48\linewidth]{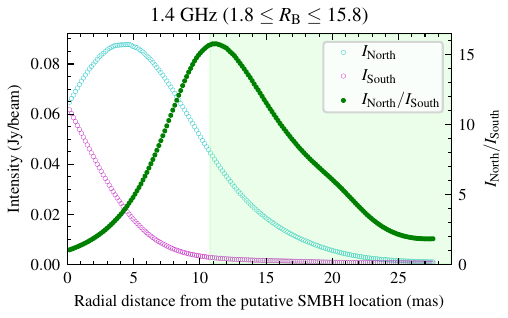}
    \includegraphics[width=0.48\linewidth]{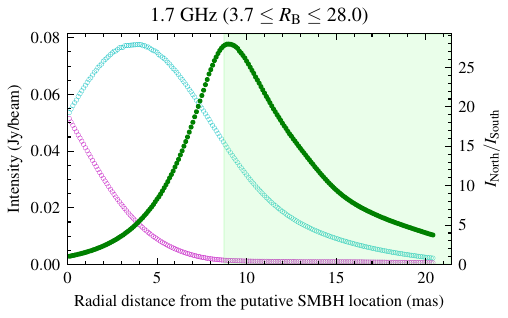}
    \includegraphics[width=0.48\linewidth]{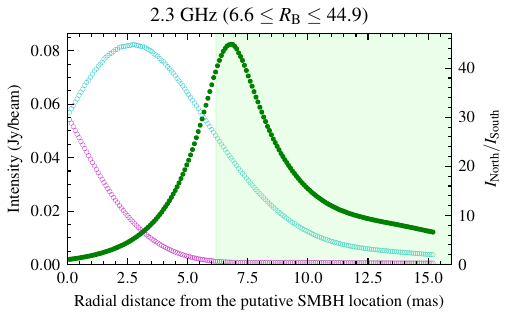}
    \includegraphics[width=0.48\linewidth]{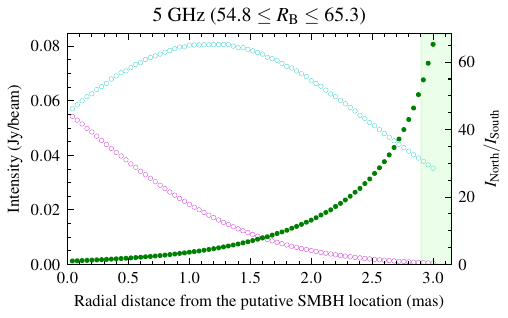}
\caption{Brightness ratio between the northern and southern jets measured at 1.4, 1.7, 2.3, and 5.0\,GHz, derived from their radial intensity profiles shown in the left panel of Figure~\ref{fig: NGC3998_intensity_profile}. Due to the limited resolution, the range of the brightness ratio is constrained by measurements within the shaded area in each image (see the text).}
\label{fig: NGC3998_BR}
\end{center}
\end{figure*}

Figure~\ref{fig: NGC3998_SIMs} displays SIMs of NGC\,3998 and the corresponding spectral distributions as a function of radial distance from the core. As shown, the northern jet exhibits optically thin spectra ($\alpha < 0$), while the southern jet, identified as the counter-jet, is dominated by optically thick, inverted spectra ($\alpha > 0$). The core, on the other hand, is characterized by self-absorbed flat spectra ($\alpha \sim 0$).

The flat-spectrum core has been observed in previous studies using observations with typical arcsecond resolutions. \citet{Hummel_1984} measured a spectral index of $\alpha = -0.17 \pm 0.05$ based on multi-frequency VLA data. Similarly, \citet{Wrobel_1984ApJ...287...41W} derived a value of $\alpha = -0.32$. More recently, \citet{Frank_2016} and \citet{Sridhar_2020A&A...634A.108S} suggested $\alpha_{\rm 1.4/5.0\,GHz} = -0.19$ and $\alpha_{\rm 1.4/4.8\,GHz} = -0.18 \pm 0.12$, respectively. Moreover, a slightly inverted spectrum, $\alpha_{\rm 147/1400\,MHz} = 0.23 \pm 0.01$, was also reported by \citet{Sridhar_2020A&A...634A.108S}. In this work, we measured the spectrum of the core on VLBI scales. The derived spectral indices, $\bar{\alpha}_{\rm 1.4/2.3\,GHz} = 0.14 \pm 0.32$, $\bar{\alpha}_{\rm 1.7/2.3\,GHz} = 0.47 \pm 0.51$, and $\bar{\alpha}_{\rm 2.3/5.0\,GHz} = -0.03 \pm 0.20$, are broadly consistent with the values reported in previous studies.


Inverted spectra of extend counter-jets have been observed in other nearby AGN. Among these, 3C\,84 is perhaps the most extensively studied, with the spectral index reaching to $\alpha > 4$ \citep{Vermeulen_1994ApJ...430L..41V,Walker_1994ApJ...430L..45W,Walker_2000ApJ...530..233W,Fujita_2017MNRAS.465L..94F,Wajima_2020ApJ...895...35W}. Additionally, inverted spectra have been detected in the approaching jet of 3C\,84, attributed to the presence of the foreground cold dense cloud \citep{Kino_2021ApJ...920L..24K,Park_2024AA...685A.115P}. The counter-jet of Centaurus\,A also shows inverted spectra with $\alpha \sim 4$ \citep{Jones_1996ApJ...466L..63J,Tingay_2001ApJ...546..210T,muller_2011A&A...530L..11M}. Furthermore, \citet{Krichbaum_1998AA...329..873K} have noted inverted spectra in Cygnus\,A. Finally, it is noted that inverted spectra are clearly observed in two prominent LLAGN, NGC\,4261 \citep{Jones_1997ApJ...484..186J,Jones_2000ApJ...534..165J,Jones_2001ApJ...553..968J,haga2015ApJ80715H,Satoh_2023PASJ...75..722S} and NGC\,1052 \citep{Kellermann_1999AAS...194.2002K,Kameno_2001PASJ...53..169K,Kameno_2003PASA...20..134K,Kameno_2023ApJ...944..156K,Vermeulen_2003AA...401..113V,Kadler_2004AA...426..481K,Baczko_2019A&A...623A..27B,Baczko_2022A&A...658A.119B,Baczko_2024A&A...692A.205B}. Theoretically, the upper limit of $\alpha$ for SSA is 2.5. Consequently, in the above cases, the highly inverted spectra observed in the counter-jets are attributed to free-free absorption (FFA) caused by the external ionized medium along the line of sight.

For NGC\,3998, we did not measure spectral indices with $\alpha>2.5$ in this study. However, the inverted spectra of the counter-jet, the evolution of its radial intensity profile (i.e., from rapid fading to a slower decline), and the frequency-dependent jet-to-counter-jet brightness ratio (see Section~\ref{sec: North/south jet brightness ratio}) suggest the possible presence of FFA in NGC\,3998. Further VLBI studies are needed to confirm this hypothesis.

\subsection{North/south jet brightness ratio} \label{sec: North/south jet brightness ratio}
Using the determined longitudinal intensity profiles shown in Figure~\ref{fig: NGC3998_intensity_profile} (left), we measured the brightness ratio of the northern and southern jets ($R_{\rm B} = I_{\rm North}/I_{\rm South}$) at each frequency, as displayed in Figure~\ref{fig: NGC3998_BR}. To mitigate blending effects between the core and the two-sided jets caused by the limited resolution, we excluded the central region having a size corresponding to twice the major-axis Full Width at Half Maximum (FWHM) of the synthesized beam (see Table~\ref{tab: NGC3998_summary_of_observation}). The determined ranges of the brightness ratio between the northern and southern jets are as follows:

\begin{align*}
\text{1.4\,GHz}: &\ 1.8 \leq R_{\rm B} \leq 15.8 \\
\text{1.7\,GHz}: &\ 3.7 \leq R_{\rm B} \leq 28.0 \\
\text{2.3\,GHz}: &\ 6.6 \leq R_{\rm B} \leq 44.9 \\
\text{5.0\,GHz}: &\ 54.8 \leq R_{\rm B} \leq 65.3 \\
\end{align*}

\begin{figure}[htbp!]
\begin{center}
    \includegraphics[width=1\linewidth]{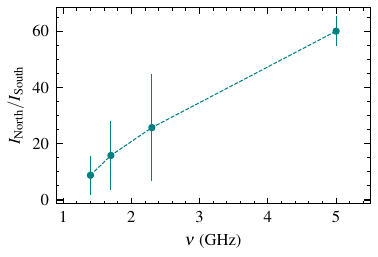}
\caption{The frequency dependence of the north/south jet brightness ratio.
\label{fig: NGC3998_BR_vs_frequency}}
\end{center}
\end{figure}  

It is noted that a peak appears in each plot, except at 5\,GHz. The peaks are basically consistent with the transition points in the radial intensity profiles of the southern jet, beyond which the southern jet fades more slowly with increasing distance (see the left panel of Figure~\ref{fig: NGC3998_intensity_profile}). Furthermore, we found that the measured brightness ratio is frequency-dependent, as shown in Figure~\ref{fig: NGC3998_BR_vs_frequency}. The similar frequency dependence of $R_{\rm B}$ has been reported in other AGN, such as Cygnus\,A \citep{Krichbaum_1998AA...329..873K}, NGC\,4261 \citep{Jones_2000ApJ...534..165J}, and NGC\,1052 \citep{Kadler_2004AA...426..481K}, which is ascribed to the FFA. Therefore, the frequency dependence of $R_{\rm B}$ observed in NGC\,3998 may be interpreted as possible evidence of FFA. However, we note that the pattern shown in Figure~\ref{fig: NGC3998_BR_vs_frequency} could result from various mechanisms. For instance, it might also be explained by a decelerating jet, where the jet-to-counter-jet brightness ratio decreases with distance, corresponding to lower frequencies \citep[e.g.,][]{park_2021ApJ...909...76P}.



\begin{deluxetable*}{ccccccc}
\tablecaption{Flux Variability of NGC\,3998 Core at 5.0\,GHz \label{tab: NGC3998_flux_variability}}
\tablehead{ \colhead{Date} & \colhead{$\nu$}  & \colhead{Array} & \colhead{$S_{\nu}$} & \colhead{$S_{\rm J1148+5254}$} & \colhead{References}\\
(yyyy/mm/dd) & (GHz) & & (mJy) &  (mJy)\\
 (1) & (2) & (3) & (4) & (5) & (6)
}
\startdata
1972/04$\sim$1972/08 & 5.0 & NRAO 91m  & $\it 73\pm13$         & ... & \citet{Sramek_1975AJ.....80..771S}\\
1979/11/30 & 4.9 & VLA  & $88\pm3\,(\star)$  & ... & \citet{Wrobel_1984ApJ...287...41W} \\
1980/06/02 & 5.0 & OEWJ & $73\pm5\,(\star)$  & ... & \citet{Hummel_1982} \\
1980/10/06 & 5.0 & OEWJ & $87\pm7\,(\star)$  & ... & \citet{Hummel_1982} \\
1980/10/08 & 5.0 & OEWJ & $76\pm8\,(\star)$  & ... & \citet{Hummel_1982} \\
1980/12/09 & 4.9 & VLA  & $82\pm2\,(\star)$  & ... & \citet{Wrobel_1984ApJ...287...41W} \\
1981/03    & 4.9 & VLA  & $76\pm2\,(\star)$  & ... & \citet{Hummel_1984}\\
1981/04/09 & 5.0 & OEWJ & $77\pm13\,(\star)$ & ... & \citet{Hummel_1982} \\
1987/10 & 4.9 & NRAO 91m  & $\it 85\pm10$ & ...& \citet{Becker_1991ApJS...75....1B,Gregory_1991ApJS...75.1011G} \\
1992/10/19 & 5.0 & VLA  & $53\pm5\,(\star)$  & ... & \citet{Laurent-Muehleisen_1997} \\
1997/06/05 & 5.0 & EVN  & $83.0\pm8.3\,(\star)$ & ... & \cite{Filho_2002} \\
2001/02/11 & 5.0 & VLBA & $86.3\pm8.6\,(\star)$   & $389.1\pm39.0$ & This work  \\
2003/01/23 & 5.0 & VLBA & $107.1\pm10.7\,(\star)$ & $371.8\pm37.2$ & This work  \\
2004/01/19 & 5.0 & VLBA & $114.3\pm11.4\,(\star)$ & $377.2\pm37.7$ & This work  \\
2004/03/19 & 5.0 & VLBA & $111.6\pm11.2\,(\star)$ & $385.4\pm38.5$ & This work  \\
2004/05/22 & 5.0 & VLBA & $115.2\pm11.5\,(\star)$ & $391.8\pm39.2$ & This work  \\
2004/07/19 & 5.0 & VLBA & $122.8\pm12.3\,(\star)$ & $388.5\pm38.9$ & This work  \\
2004/09/16 & 5.0 & VLBA & $124.9\pm12.5\,(\star)$ & $391.0\pm39.1$ & This work  \\
2004/11/20 & 5.0 & VLBA & $136.0\pm13.6\,(\star)$ & $397.5\pm39.8$ & This work  \\
2006/02/24 & 4.9 & VLA  & $\it 302.0\pm30.2$   & ... & \citet{Kharb_2012AJ....143...78K}\\
2006/05/27 & 4.8 & VLBA & $270.1\pm27.0\,(\star)$ & ... & \citet{Helmboldt_2007ApJ...658..203H}  \\
2015/06/20 & 4.9 & WSRT & $118\pm6\,(\star)$ & ... & \citet{Frank_2016}  \\
\enddata
\tablecomments{
Column\,(1): observing date. 
Column\,(2): frequency.
Column\,(3): array used in observation. ``NRAO 91m'' refers to the NRAO Green Bank 300-foot telescope, and ``OEWJ'' denotes a VLBI array consisting of Onsala, Effelsberg, Westerbork, and Jodrell Bank. EVN is the European VLBI Network, and WSRT is the Westerbork Synthesis Radio Telescope.
Column\,(4): flux density of NGC\,3998. The stars in brackets indicate the flux density of the core, while italics denote the total flux. 
Column\,(5): the total flux density of the calibrator J1148+5254.
Column\,(6): references.
}
\end{deluxetable*}

\begin{figure*}[htbp!]
\vspace{-0.8cm}
\begin{center}
    \begin{minipage}{0.48\linewidth}
        \centering
        \includegraphics[width=1\linewidth]{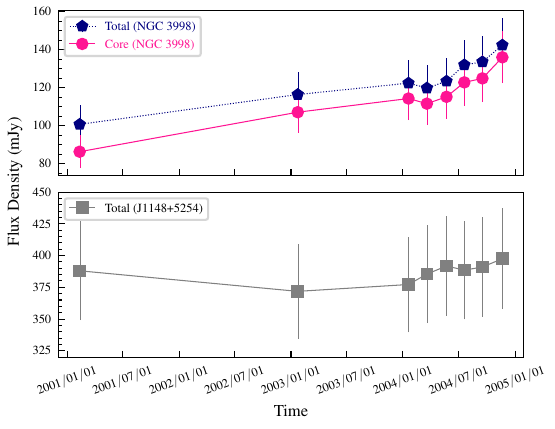}
    \end{minipage}
    \begin{minipage}{0.5\linewidth}
        \centering
        \includegraphics[width=1\linewidth]{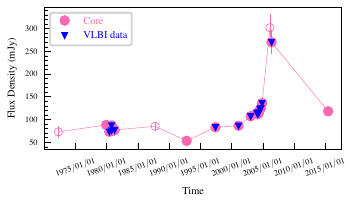}
    \end{minipage}%
    \caption{Left: flux density variability of NGC\,3998 core at 5.0\,GHz from early 2001 to late 2004 (top panel). As a control of amplitude calibration, the total flux density of the calibrator J1148+5254 is shown in the bottom panel. Right: the long-term 5.0\,GHz radio light curve of NGC\,3998. The open circles represent the total flux density, while the filled circles correspond to the core flux density. Measurements from VLBI observations are highlighted with blue triangles (see Table~\ref{tab: NGC3998_flux_variability}).
    \label{fig: NGC3998_variability} }
\end{center}
\end{figure*}

\subsection{Flux density variability} \label{sec: Flux density variability}
The total flux density observed at 5\,GHz, as listed in Table~\ref{tab: NGC3998_summary_of_observation}, suggests that NGC\,3998 shows variability between February 2001 and November 2004. As suggested by \citet{Frank_2016}, this variability is associated with the core. Therefore, we performed model fitting on the multi-epoch 5\,GHz dataset to derive the flux density of the VLBI core. As summarized in Table~\ref{tab: NGC3998_flux_variability} and shown in Figure~\ref{fig: NGC3998_variability}, these results indicate a $\sim$60\% increase in the core flux density from early 2001 to late 2004. Moreover, we also show the flux density of the calibrator J1148+5254 as a control of amplitude calibration, which shows no significant flux variation ($<$10\%; see Table~\ref{tab: NGC3998_flux_variability}).

To further extend our analysis, we compiled the 5\,GHz flux density of NGC\,3998 from the literature, spanning over 40 years (see Table~\ref{tab: NGC3998_flux_variability}). As presented in the right panel of Figure~\ref{fig: NGC3998_variability}, the core shows a relatively stable flux density from the 1970s through the 1990s. However, from 2000 onward, its flux density starts to increase, which is followed by a relatively sharp rise --- exceeding a factor of two --- between late 2004 and early 2006. This significant flux increase is likely associated with a new jet ejection event, potentially triggered by a fueling episode in the nuclear region, as discussed in detail by \citet{Frank_2016}.

Interestingly, NGC\,3998 also shows long-term variability in the X-ray band.  As suggested by \citet{Ptak_2004ApJ...606..173P}, the 0.5–2.0\,keV flux density increased from $\sim (3.7\pm0.7)\times10^{-12}{\,\rm erg\,s^{-1}\,cm^{-2}}$ in 1980 to $\sim (6.9\pm1.3)\times10^{-12}{\,\rm erg\,s^{-1}\,cm^{-2}}$ in 2001. Additionally, \citet{Younes_2011A&A...530A.149Y} reported a slightly decreased flux density of $\sim 5.6_{-0.1}^{+0.3}\times10^{-12}{\,\rm erg\,s^{-1}\,cm^{-2}}$ in July 2006 at the same energy band, immediately following the radio peak. On the other hand, from Figure~4 in \citet{Ptak_2004ApJ...606..173P}, we noted a gradual increase of the X-ray flux density by a factor of approximately 1.8 from 1991 to 2001. Notably, the radio flux density also increased by a comparable factor of about 1.6 from 1992 to 2001 (see Table~\ref{tab: NGC3998_flux_variability}). Therefore, a correlation may exist between the radio and X-ray variability. Detailed studies, involving the collection of more historical data in both radio and X-ray regimes, would be essential to determine the exact correlation and to better understand the role of the jet in the X-ray emission process \citep[e.g.,][]{Ptak_2004ApJ...606..173P,Pian_2010MNRAS.401..677P}.

\section{Summary} \label{sec: summary}
Using archival VLBA data, we have presented the multi-frequency, multi-epoch VLBI study of the jet in NGC\,3998. Our primary findings are as follows:

\begin{enumerate}

\item We detected symmetric two-sided jets on VLBI scales, with a total extent of $\sim 80$\,mas (or $\sim 5.3$\,pc). The position angle of the pc-scale jet (observed in 2001) differs by about $26^\circ$--$30^\circ$ from that of the kpc-scale jet (observed in 2015), which may be indicative of jet precession. 
We also measured frequency-dependent core shifts at frequencies between 1.4 and 5 GHz. We note that further VLBI observations are essential to better constrain the core-shift relationship and more accurately pinpoint the location of the central SMBH. Finally, based on the core shift, radial intensity profile of jet, and north/south jet brightness ratio, we identified the northern jet as the approaching jet and the southern jet as the counter-jet.

\item Our measured radial intensity profiles of the twin jets suggest a change in the counter-jet emission from rapid fading to a slower decline. Spectral analysis shows that the approaching jet exhibits an optically thin spectrum ($\alpha < 0$), while the counter-jet is dominated by an optically thick, inverted spectrum ($\alpha > 0$). Moreover, we find that the north/south jet brightness ratio is frequency-dependent. These findings may hint at the presence of FFA in NGC\,3998. Further observations are needed to confirm this hypothesis.

\item Multi-frequency and multi-epoch observations revealed a compact, flat-spectrum, and variable core on VLBI scales. The long-term light curve indicates a relatively stable flux density from the 1970s through the 1990s, followed by a gradual increase starting in 2001, peaking in early 2006. The significant flux increase is likely associated with a new jet ejection event. Moreover, there may be a correlation between the radio and X-ray variability, which needs further confirmation through more extensive data.


\end{enumerate}

\begin{acknowledgments}
We sincerely thank the anonymous referee for the valuable and constructive comments, which helped to greatly improve the quality of the manuscript. This work was supported by the CAS `Light of West China' Program (grant No. 2021-XBQNXZ-005) and the National SKA Program of China (grant No. 2022SKA0120102). XY acknowledges the support from the 2025 outstanding post-doctoral grant of Xinjiang Uygur Autonomous Region. LC acknowledges the support from the Tianshan Talent Training Program (grant No. 2023TSYCCX0099). XY and LCH acknowledge the support from the Xinjiang Tianchi Talent Program. LCH was supported by the National Key R\&D Program of China (2022YFF0503401), the National Science Foundation of China (11991052, 12233001), and the China Manned Space Project (CMS-CSST-2021-A04, CMS-CSST-2021-A06). This work was also partly supported by the Urumqi Nanshan Astronomy and Deep Space Exploration Observation and Research Station of Xinjiang (XJYWZ2303). The Very Long Baseline Array is operated by the National Radio Astronomy Observatory, which is a facility of the National Science Foundation operated under cooperative agreement by Associated Universities, Inc.
\end{acknowledgments}

\bibliographystyle{aasjournal}
\bibliography{references}

\appendix
\section{Multi-epoch 5\,GHz images of NGC\,3998} \label{Appendix: The multi-epoch 5 GHz CLEAN images}
\begin{figure*}[htbp!]
\begin{center}
    \includegraphics[width=0.32\linewidth]{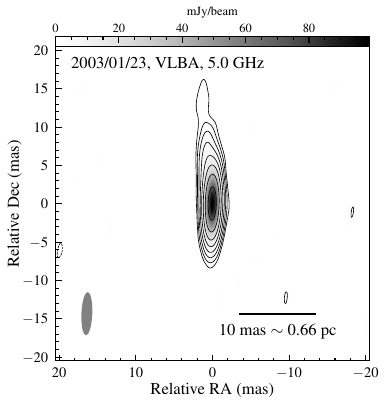}
    \includegraphics[width=0.32\linewidth]{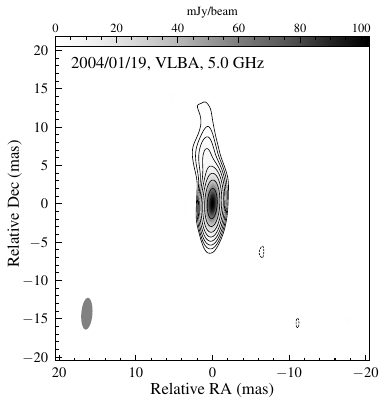}
    \includegraphics[width=0.32\linewidth]{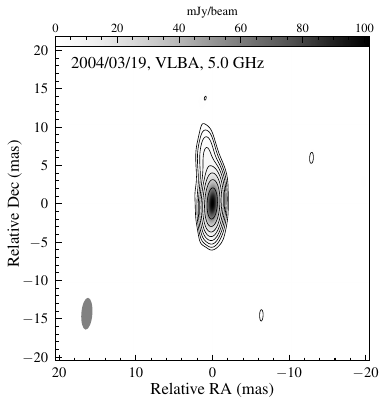}
    \includegraphics[width=0.32\linewidth]{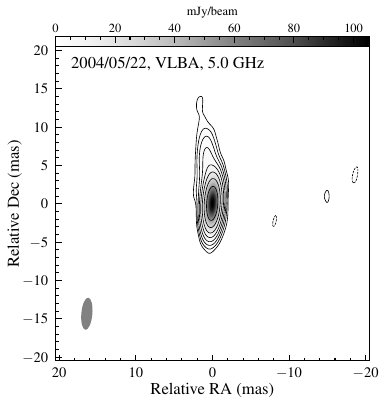}
    \includegraphics[width=0.32\linewidth]{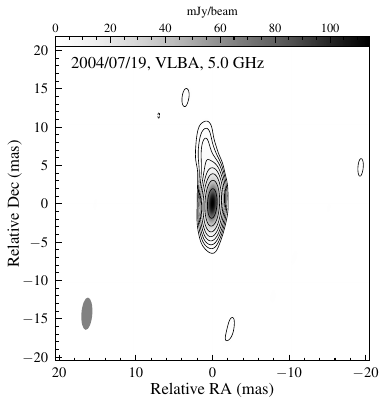}
    \includegraphics[width=0.32\linewidth]{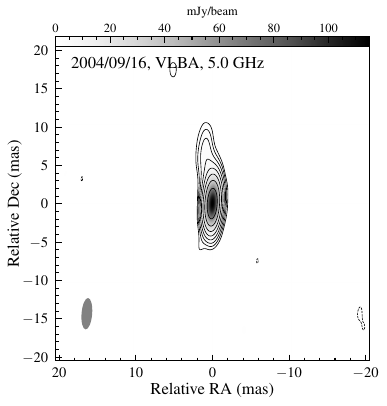}
    \includegraphics[width=0.32\linewidth]{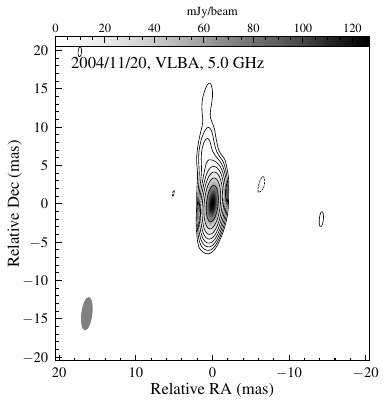}
\caption{Multi-epoch 5\,GHz images of NGC\,3998 obtained with the natural weighting scheme. The synthesized beam is shown in the bottom-left corner of each image. Contours are plotted at $-1$, $1$, $2$, $4$, ..., times the $3\sigma$ noise level of each image (see Table~\ref{tab: NGC3998_summary_of_observation}), increasing by a factor of 2.
\label{fig: NGC3998_CLEAN_images_5GHz}}
\end{center}
\end{figure*}

\section{Calibrator J1148+5254} \label{Appendix: Calibrator J1148+5254}
\begin{figure}[htbp!]
\begin{center}
    \includegraphics[width=0.45\linewidth]{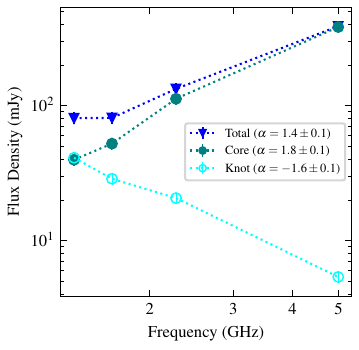}
\caption{Radio spectra of J1148+5254.
\label{fig: J1148+5254_spectra}}
\end{center}
\end{figure}

J1148+5254 is a quasar with a redshift of \(z = 0.373\) \citep[e.g.,][]{2024AJ....168...58D}. As shown in Figure~\ref{fig: J1148+5254_PRed_CLEAN}, this source exhibits a ``core + knot" structure. Using the self-calibrated images, we modeled the source structure at each frequency. The model-fitted flux densities of the core and knot are listed in Table~\ref{tab: Flux density of J1148+5254}. The radio spectra of J1148+5254 are shown in Figure~\ref{fig: J1148+5254_spectra}, revealing an optically thick core and an optically thin knot (cf. Figure~\ref{fig: J1148+5254_PRed_CLEAN}).

\begin{deluxetable}{cccc}[htbp!]
\tablecaption{Flux Density of J1148+5254 \label{tab: Flux density of J1148+5254}}
\tablehead{ \colhead{$\nu$} & \colhead{$S_{\rm tot}$} & \colhead{$S_{\rm core}$} & \colhead{$S_{\rm knot}$}\\
(GHz) & (mJy)  & (mJy)  & (mJy)   \\
(1) & (2)& (3)& (4)}
\startdata
1.4 & $81.0\pm8.1$   & $39.8\pm4.0$   & $41.2\pm4.1$  \\
1.7 & $81.2\pm8.1$   & $52.4\pm5.2$   & $28.8\pm2.9$  \\
2.3 & $133.2\pm13.3$ & $112.5\pm11.3$ & $20.7\pm2.1$  \\
5.0 & $387.7\pm38.8$ & $382.3\pm38.2$ & $5.4\pm0.5$  \\
\enddata
\tablecomments{
Column\,(1): frequency.
Column\,(2): total flux density of J1148+5254.
Columns\,(3)-(4): flux densities of the core and knot, respectively. The flux density error is assumed to be 10\%.
}
\end{deluxetable}

\end{document}